\documentclass[doublecol,figures]{epl2} 
\usepackage{graphicx}
\usepackage{amsmath,amssymb}
\usepackage{bm}

\title{A note on the possibility of classical orbital diamagnetism for an unbounded system --the Bohr-van Leeuwen Theorem} 

\shorttitle{Classical diamagnetism}

\author{K. Vijay Kumar \inst{1}\thanks{email: \email{vijayk@physics.iisc.ernet.in}} 
		\and N. Kumar \inst{2,3}\thanks{email: \email{nkumar@rri.res.in} }}
\shortauthor{K. Vijay Kumar \etal}

\institute{                    
  \inst{1} CCMT, Dept. of Physics, Indian Institute of Science, Bangalore, 
						India 560012 \\
	\inst{2} Raman Research Institute, Bangalore, India 560080\\
	\inst{3} Jawaharlal Nehru Centre for Advanced Scientific Research, 
					Bangalore, India 560064
}

\abstract{
Recently [EPL, \textbf{86}, (2009) 17001], we had simulated the classical Langevin dynamics of a charged
particle on the surface of a sphere in the presence of an externally
applied  magnetic field, and found a finite value for the orbital
diamagnetic moment in the long-time limit. This result is surprising in
that it seems to violate the classic  Bohr-van Leeuwen Theorem on the
absence of classical diamanetism. It was indeed questioned by some workers 
[EPL, \textbf{89}, (2010) 37001] who
verified that the Fokker-Planck (FP) equation derived from  our Langevin
equation, was satisfied by the classical canonical density in the steady state, obtained by setting $\partial / \partial t = 0$ in the FP equation. 
Inasmuch as the canonical density does not contain the magnetic field,
they concluded that the diamagnetic moment must be zero. 
The purpose of this note is to show that
this argument and the conclusion are invalid -- instead of setting 
$\partial / \partial t = 0$, 
one must first obtain the fundamental time-dependent solution for the FP equation, 
and then calculate the expectation value of the diamagnetic moment, and finally
consider its long-time limit (i.e., $t \to \infty$). This would indeed 
correspond to our numerical simulation of the dynamics. That this is indeed
so is shown by considering the simpler analytically solvable problem, namely that  for an unbounded plane for which the above procedure can be carried
out exactly. We then find that the limiting value for the
expectation of the  diamagnetic moment is indeed non-zero, and yet
the steady-state FP equation  obtained by setting $\partial / \partial t = 0$ 
is satisfied by the canonical density. Admittedly, the exact analytical solution 
for the sphere is not available. But, the exact solution obtained for the case of the
unbounded 2D-plane illustrates our point all right. We also present some further new results for other finite but unbounded surfaces such the ellipsoids of revolution.
}

\begin{document}

\maketitle

%% Introduction
 Absence of classical orbital diamagnetism -- the Bohr-van Leeuwen (BvL) theorem \cite{vanLeeuwen,BohrThesis} has been a surprise of physics \cite{vanVleck,Peierls}. 
 This is so because the
amperean current associated with the closed classical orbits of a
charged particle under the influence of the Lorentz force is expected
to subtend a non-zero moment that should be diamagnetic -- the
Lenz's law. Bohr had given a physical argument for this null result in
terms of the cancellation of the diamagnetic moment of the completed
orbits of the charged particle away from the boundary  by the
paramagnetic moment due to the incomplete orbits skipping the boundary
in the opposite sense. This surprising cancellation had motivated us in
a previous work \cite{KumarVijay} to examine the case of a charged particle moving on a
finite but unbounded closed surface, namely, the surface of a
sphere, that naturally has no boundary. To our surprise, the long-time
average of the orbital motion indeed gave a finite diamagnetic moment. 
The motion was treated through the stochastic Langevin dynamics that
is expected to give equilibrium in the long-time limit -- the
Einsteinian approach to statistical mechanics \cite{Kadanoff}. The unbounded nature of the sphere ensured avoided cancellation. The above result obtained by us has been recently questioned by some workers \cite{PradhanSeifert} who
verified that the Fokker-Planck (FP) equation derived from  our Langevin
dynamics was satisfied by the classical canonical density in the steady state 
obtained by setting $\partial / \partial t = 0$ in the FP equation. 
Inasmuch as the canonical density does not contain the magnetic field,
they concluded that the diamagnetic moment is zero. 
The purpose of this note is to show that
this conclusion is invalid -- instead of setting $\partial / \partial t = 0$, 
one must first obtain the fundamental time-dependent solution for the FP equation, 
then calculate the expectation value of the diamagnetic moment, and finally
consider its long-time limit (i.e., $t \to \infty$) -- this would indeed 
correspond to our numerical simulation of the dynamics. That this is indeed
so is shown by considering the simpler but, analytically solvable problem, namely that  for an unbounded plane for which the above procedure can be carried
out exactly. We then find that the limiting value for the
expectation of the  diamagnetic moment is indeed non-zero, and yet
the steady-state FP equation obtained by setting $\partial / \partial t = 0$ 
is satisfied by the canonical density. Admittedly, the exact analytical solution 
for the sphere is not available. But, the exact solution for the case of the
unbounded 2D-plane illustrates our point all right. We also present some further new results for other finite but unbounded surfaces such as the ellipsoids of revolution.

Let us consider a charged particle (of charge $-|e|$ and mass $m$) moving with a velocity 
${\bf u} = (u_x, u_y)$ in a two-dimensional unbounded plane in the presence of a uniform magnetic field ${\bf B}$ perpendicular to the plane and under the influence of a stochastic force and the  concomittant dissipation as in the classical Langevin equation
\begin{equation}
\dot{\bf u} = -\gamma {\bf u} - \frac{|e|}{m c} {\bf u} \times {\bf B} + f(t),
\end{equation}
where $\gamma$ is the frictional relaxation rate, $f(t)$ is a Gaussian white noise satisfying
\begin{equation}
\langle f_i(t) f_j(t') \rangle = \frac{2 \gamma k_B T}{m} \delta_{ij} \delta(t-t').
\end{equation}
The corresponding Fokker-Planck (FP) equation can now be written as \cite{Romero}
\begin{eqnarray}
\frac{\partial P}{\partial t} 
&=& - u_x \frac{\partial P}{\partial x} 
- u_y \frac{\partial P}{\partial y}
+ \frac{\partial \big[ (\gamma u_x + \omega_c u_y)P \big]}{\partial u_x}
\nonumber \\
&& + \frac{\partial \big[ (\gamma u_y - \omega_c u_x)P \big]}{\partial u_y}
+ \frac{\gamma k_BT}{m} \Big(\frac{\partial^2 P}{\partial u_x^2} 
+ \frac{\partial^2 P}{\partial u_y^2}\Big) \quad
\label{eq:fpeqn}
\end{eqnarray}
for the \textit{normalized} transition probability density 
$P({\bf r},{\bf u}, t | {\bf r}_o,{\bf u}_o)$ that describes the stochastic evolution of the system, where ${\bf r} = (x,y)$ is the position vector of the particle. Following the method due originally to Chandrasekhar \cite{Chandrasekhar}, the exact analytical solution of the above FP equation can be readily written down \cite{Romero} as
\begin{eqnarray}
&& 
%\hspace{-0.5cm} 
P
%(x,y,u_x,u_y,t) 
= \mathcal{N} \exp\left\{ -\frac{2 \pi^2 \mathcal{N}}{C_1} \, 
\Big( F |{\bf u}|^2 - 2 H {\bf r} \cdot {\bf u}  \right.
\nonumber \\
&& \qquad \left.
+ 2 H \, \dfrac{\omega_c}{\gamma} \, ({\bf u} \times {\bf r})_z + C_1 G |{\bf r}|^2 \Big)
\right\}
\nonumber \\
&& \quad = \mathcal{N} \exp\left\{ 
-\frac{2 \pi^2 \mathcal{N}}{C_1} \, \Big( F (u_x^2 + u_y^2) - 2 H (x u_x + y u_y) \right.
\nonumber \\
&& \qquad \left.
+ 2 H \, \dfrac{\omega_c}{\gamma} \, (y u_x - x u_y) + C_1 G (x^2 + y^2) \Big)
\right\},
\label{eq:fpsoln}
\end{eqnarray}
where without loss of generality we have set the initial conditions 
${\bf r}_o = {\bf u}_o= 0$, i.e., a particle starting from rest at the origin. In the above equation $\omega_c = |e|B/mc$ is the cyclotron frequency and
\begin{eqnarray}
\mathcal{N} &=& \frac{C_1}{4 \pi^2 (FG-H^2)},
\\
C_1 &=& \dfrac{\gamma^2 + \omega_c^2}{\gamma^2},
\\
F &=& \dfrac{k_B T}{m \gamma^2} (2 \gamma t -3 + 4 e^{-\gamma t} - e^{-2 \gamma t}),
\\
G &=& \dfrac{k_B T}{m} (1 - e^{-2 \gamma t}),
\\
H &=& \dfrac{k_B T}{m \gamma} (1 - e^{-2 \gamma t})^2.
\end{eqnarray}
Note that equation (\ref{eq:fpsoln}) is not invariant under field reversal 
${\bf B} \to -{\bf B}$.

Using the above solution for the FP equation, we can readily calculate the expectation of the orbital magnetic moment
\begin{equation}
m = -\frac{|e|}{2c}  \, (x u_y - y u_x)
\end{equation}
as
\begin{eqnarray}
&& \hspace{-1cm} \langle m(t) \rangle = -\frac{|e|}{2c} 
\int_{-\infty}^{\infty} dx \int_{-\infty}^{\infty} dy
\int_{-\infty}^{\infty} du_x \int_{-\infty}^{\infty} du_y (x u_y - y u_x) P
\nonumber \\
&=& -\frac{|e|}{2c} \frac{C_1 \gamma}{4 \pi^2 H \omega_c}  \frac{\partial }{\partial \alpha} 
\Big[
\int_{-\infty}^{\infty} dx \int_{-\infty}^{\infty} dy
\int_{-\infty}^{\infty} du_x \int_{-\infty}^{\infty} du_y
\nonumber \\
&& \exp\left\{ 
-\frac{2 \pi^2 \mathcal{N}}{C_1} \, \Big( F (u_x^2 + u_y^2) - 2 H (x u_x + y u_y) \right.
\nonumber \\
&& \left.
+ 2 H \, \dfrac{\omega_c}{\gamma} \alpha \, (y u_x - x u_y) + C_1 G (x^2 + y^2) \Big)
\right\} \Big] \Bigg\vert_{\alpha=1} \quad
\nonumber \\
&=& -\frac{|e|}{mc} \, \frac{k_B T \, \omega_c}{(\gamma^2 +\omega_c^2)} \, (1 - e^{-2\gamma t})^2.
\label{eq:magtime}
\end{eqnarray}
Note that the expression for $\langle m \rangle$ is odd in ${\bf B}$ (or $\omega_c$). In the long-time limit
\begin{equation}
\lim_{t \to \infty} \langle m(t) \rangle  = 
-\frac{|e|}{mc} \, \frac{k_B T \, \omega_c}{(\gamma^2 +\omega_c^2)}
\neq 0.
\label{eq:maglimit}
\end{equation}
(This is precisely the expression given in equation (8) of \cite{KumarJayannavar}). It may also be noted from equation (\ref{eq:magtime}) that the time-scale involved in the growth of the magnetic moment from zero to full value is order $\gamma^{-1}$ which is a microscopic quantity. However, the diamagnetic moment gets distributed over the whole space on a much longer time-scale.

% Discussion
We now return to the main point of our contention. First, following \cite{PradhanSeifert}, 
we verify that the classical (canonical) Boltzmann equilibrium distribution namely
\begin{equation}
P_{can} \sim \exp\left\{-\frac{m}{2k_BT} (u_x^2 + u_y^2) \right\}
\end{equation}
when substituted on the right hand side of equation (\ref{eq:fpeqn}) does give zero. This corresponds to a statement made in \cite{PradhanSeifert} that the classical canonical distribution satisfies their FP equation with $\partial / \partial t = 0$, i.e., steady state. The exact time-dependent solution, however, leads to a non-zero diamagnetic moment in the limit $t \to \infty$ as seen from equations (\ref{eq:magtime}) and  (\ref{eq:maglimit}). This clearly shows that the arguement of \cite{PradhanSeifert} leading to the conclusion that classical diamagnetism is zero, is invalid. 

To reemphasize this point further in the context of the motion on a sphere, we would like to note that even the Liouville equation of \cite{PradhanSeifert} for a particle moving on the surface of a sphere has other solutions too \emph{and with non-zero currents}. In fact, using \textit{Mathematica} \cite{Wolfram}, we found that (notation as in \cite{PradhanSeifert})
\begin{equation}
\rho_{st} = \sin\theta \,\, \mathcal{F} (c_1, c_2),
\label{eq:liouvillesoln}
\end{equation}
where $c_1$ and $c_2$ are given by
\begin{eqnarray}
c_1 &=& v_{\phi} \, \sin\theta - \frac{1}{2} a \, \omega_c \cos^2\theta,
\\
\nonumber \\
c_2 &=& \frac{1}{64} \big[ 4 ( 8 v_{\theta}^2 + 4 v_{\phi}^2 + a^2 \omega_c^2) + 
	4 (4 v_{\phi}^2 - a^2 \omega_c^2 )  \, \cos 2\theta
\nonumber \\
&& 
	+ a \, \omega_c (32 v_{\phi} \sin \theta \cos^2\theta - a \, \omega_c \cos 4\theta ) \big],
\end{eqnarray}
and $\mathcal{F}$ an \emph{arbitrary function} of its two arguments $c_1$ and $c_2$, satisfies equation (8) of \cite{PradhanSeifert}. And yet the solutions, equation (\ref{eq:liouvillesoln}), too have a finite magnetic moment inasmuch as they are not invariant under the inversion of field ${\bf B} \to -{\bf B}$.

%% Ring
The cases of unbounded systems studied above
clearly point to the role of the boundary in the context of the BvL theorem in
that the absence of the boundary gives the avoided cancellation, and
hence the non-zero  orbital diamagnetism. In this connection it will be apt to
clarify the case of a circular ring which is also an unbounded
system, and yet the associated orbital diamgnetism is readily seen to
vanish \cite{KaplanMahanti}. Is this a counter-example? The answer is a definite no. The
vanishing of the orbital diamagnetic moment in this case is due simply to
the fact that here the only allowed motion is
necessarily along the tangent at any  point on the ring. The Lorentz force, therefore, 
must act in a plane normal to this tangent. The latter can have no effect whatsoever on the
tangential velocity because of the holonomic radial (rigid)
constraint that balances all radial forces. Thus, the magnetic field
generates no orbital response for this constrained system. 
This point has also been noted by \cite{PradhanSeifert}.  Clearly, for any non-zero effect of the magnetic field on the orbital motion at all, the tangent space must be at least two dimensional, as for the case of a sphere for example. 
Thus, for the case of a ring (or for that matter, for any non-planar loop in
general) the magnetic field is irrelevant, and these cases  do not constitute valid counter examples.

%% ellipsoid
The surface of a sphere is, of course, rather special in that it has a constant curvature, and the avoided cancellation may well be a consequence of this, it may be
argued. This has further motivated us to consider in the present work
the more general case of a closed surface of non-constant curvature, namely, an ellipsoid of revolution. We consider a charged particle of charge $-|e|$, mass $m$ moving on the surface of an ellipsoid of revolution around the $z-$axis with semi-major axis $a$ and semi-minor axis 
$b = a \, \epsilon$, where $\epsilon$ is the ellipticity. The position of the particle can be described by the polar angle $\theta$ and the azimuthal angle $\phi$. The Langevin equations of the particle in the tangential ($\theta-$like) and the azimuthal direction 
are given by
\begin{eqnarray}
&& \hspace{-1cm}  \frac{(\epsilon^2 \sin^2\theta + \cos^2\theta) \, \ddot{\theta} + 
	\sin\theta \cos\theta \, \big[ (\epsilon^2-1) \, \dot{\theta}^2 - \dot{\phi}^2 \big]} 
	{\sqrt{\epsilon^2 \sin^2\theta + \cos^2\theta}}
\nonumber \\
&& \hspace{-1cm}  \quad
= -\frac{\omega_c}{\gamma} \, \frac{\sin\theta \, \cos\theta}
		{\sqrt{\epsilon^2 \sin^2\theta + \cos^2\theta}} \, \,  \dot{\phi} 
\nonumber \\
&& \hspace{-1cm}  \qquad
   - \sqrt{\epsilon^2 \sin^2\theta + \cos^2\theta} \, \dot{\theta} 
   + \sqrt{\eta} \, \,  f_{\theta}(t),
\\
\nonumber \\
&& \hspace{-1cm} 
\sin\theta \, \ddot{\phi} + 2 \cos\theta \, \dot{\theta} \, \dot{\phi}
= \frac{\omega_c}{\gamma} \, \cos\theta \, \,  \dot{\theta} 
 -  \sin\theta  \,  \, \dot{\phi} + \sqrt{\eta} \, \,  f_{\phi}(t)
\end{eqnarray}
where $\eta = 2 k_BT / (ma^2 \gamma^2)$. The notations above are again the same as in our earlier paper \cite{KumarVijay}. The area normalized magnetic moment
\begin{equation}
\langle \mu_{\epsilon} \rangle = \frac{-\langle \sin^2\theta \, \dot{\phi} \rangle}
	{\dfrac{1}{2} + \dfrac{\epsilon^2}{2\sqrt{1-\epsilon^2}} 
	\sinh^{-1} \Big[ \dfrac{\sqrt{1-\epsilon^2}}{\epsilon}  \Big]}.
\label{eq:magmom}
\end{equation}
As is clear from Fig. 1, it gave, to our pleasant surprise, 
a diamagnetic moment that now depended systematically on the
ellipticity parameter $\epsilon$. This case too confirms the avoided cancellation responsible for our orbital diamagnetism.

\begin{figure}[t]
\onefigure[width=\columnwidth]{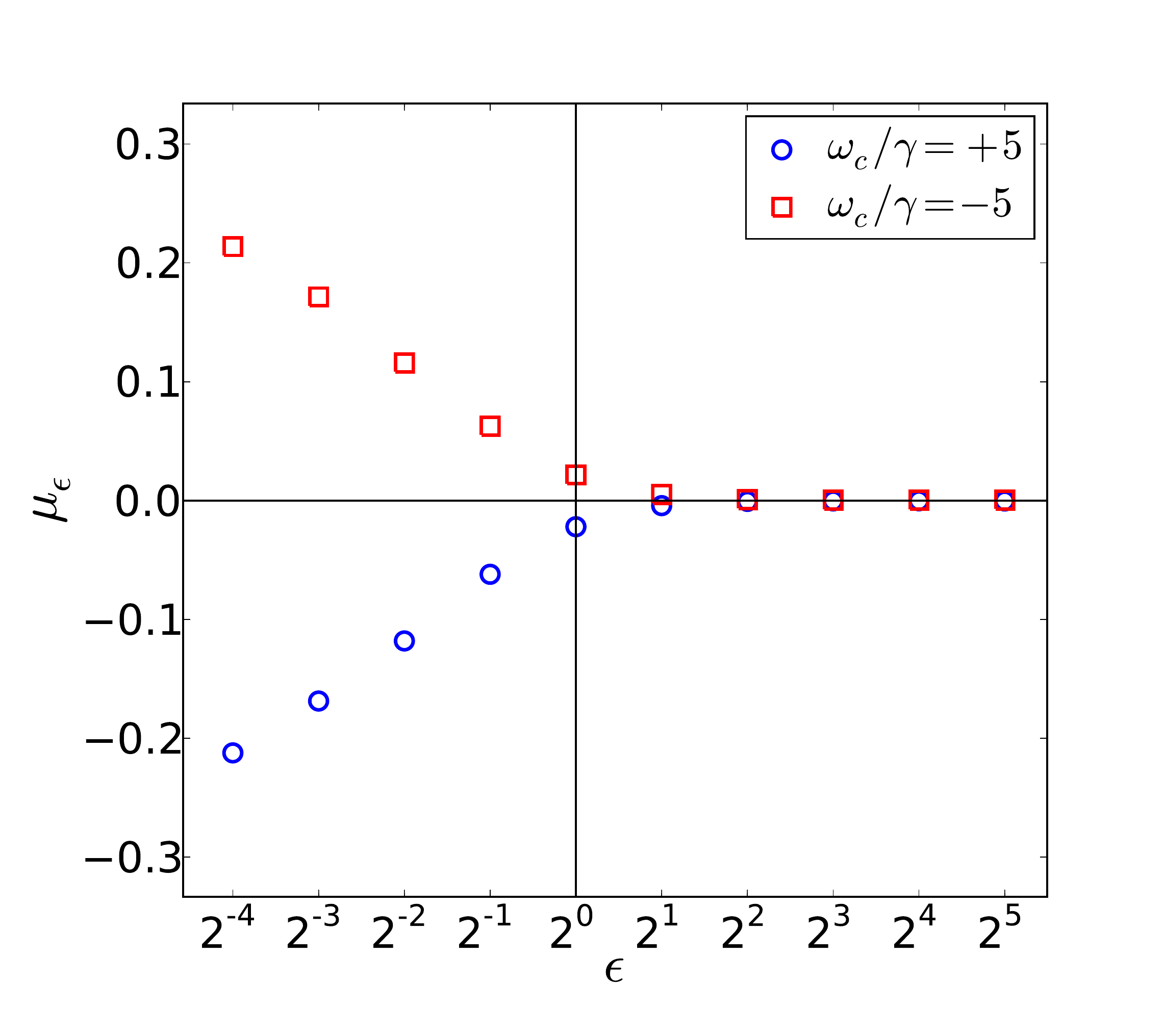}
\caption{{\small (Color online)} Plot of the area normalized magnetic moment, 
$\mu_{\epsilon}$ eqn. (\ref{eq:magmom}), vs. the ellipticity $\epsilon$ for an ellipsoid 
of revolution. Here $\omega_{c}/\gamma = \pm 5.0$ and $\eta = 1.0$.} 
\label{fig:distribution}
\end{figure}

% CONCLUSION
 In conclusion, the classical Langevin dynamics of a charged particle
moving in a magnetic field on an unbounded surface is found to
give non-zero orbital diamagnetism as found in our earlier paper \cite{KumarVijay}. 
We have successfully provided a rebuttal to the objection raised in \cite{PradhanSeifert} and \cite{KaplanMahanti}. Further, we have generalized our earlier results to surfaces of 
nonconstant curvature, namely an ellipsoid of revolution. The results clearly suppport the idea of avoided cancellation. The diamagnetic moment is found to depend systematically on the ellipticity.

% Thanks
\begin{acknowledgements}
We thank \textsc{T.~A.~Kaplan}, \textsc{P.~Pradhan} for correspondence and \textsc{S.~Ramaswamy} for many fruitful discussions.
\end{acknowledgements}

% REFERENCES

%\bibliography{diamagnetism_ref}{}
%\bibliographystyle{eplbib}

\end{document}